\documentclass[UTF8]{article}
\usepackage{authblk}

\title{Probing the gravitational wave background from 
	cosmic strings with Alternative LISA-TAIJI network}
\author{Bo-Rui Wang$^1$,
	Jin Li$^{1,}$$^{2,}$\thanks{Jin Li: cqujinli1983@cqu.edu.cn}, He Wang$^{3,}$$^4$}
\affil{\begin{small}
		$^{1}$ College of Physics, Chongqing University, Chongqing 401331, China
	\end{small}}
\affil{\begin{small}$^{2}$ Department of Physics and Chongqing Key Laboratory for Strongly Coupled Physics, Chongqing University, Chongqing 401331, China\end{small}}
\affil{\begin{small}$^{3}$ International Centre for Theoretical Physics Asia-Pacific, University of Chinese Academy of Sciences, 100190 Beijing, China\end{small}}
\affil{\begin{small}$^{4}$ Taiji Laboratory for Gravitational Wave Universe, University of Chinese Academy of Sciences, 100049 Beijing, China\end{small}}

\date{\today}
\usepackage{graphicx,indentfirst,amsmath,cite,enumerate,pifont}
\usepackage[section]{placeins}
\begin{document}
	\maketitle
	\begin{abstract}
		As one of the detection targets of all gravitational wave detectors at present, stochastic gravitational wave background (SGWB) provides us an important way to understand the evolution of our universe. In this paper, we explore the feasibility of detecting the SGWB generated by the loops, which arose throughout the cosmological evolution of the cosmic string network, by individual space detectors (e.g.~LISA, TAIJI) and joint space detectors (LISA-TAIJI). For joint detectors, we choose three different configurations of TAIJI (e.g.~TAIJIm, TAIJIp, TAIJIc) to form the LISA-TAIJI networks. And we investigate the performance of them to detect the SGWB. Though comparing the power-law sensitivity (PLS) curves of individual space detectors and joint detectors with energy density spectrum of SGWB. We find that LISA-TAIJIc has the best sensitivity for detecting the SGWB from cosmic string loops and is promising to further constrains the tension of cosmic sting $G\mu=O(10^{-17})$.
	\end{abstract}
	\section{Introduction}
	Gravitational waves (GWs), which are generated from violent movement and change of matter and energy, carry very important information about their sources. At the beginning of the universe, there was full of dense matter, so that the gravitational waves generated by collisions between particles were immediately absorbed by other particles. In the inflationary stage of the rapid expansion of the universe, the density of the universe dropped suddenly, and the released gravitational waves were no longer absorbed. Since then, those primitive disturbances had spread in the space around us to form a stochastic gravitational wave background (SGWB). The SGWB is a superposition of substantial incoherent GWs, including cosmological sources such as phase transitions\cite{name48,name49,name50,name51}, cosmic strings\cite{name10,name52,name53} and inflation models\cite{name54,name55}. SGWB has an extremely wide frequency band ($10^{-18}$--$10^{10}$Hz)\cite{name66,name67}. So that, the waves sources of all detectors contain SGWB. NANOGrav reported a stochastic process from the 12.5-$yr$ data set\cite{name57}, that gives us a promising expectation to detect SGWB.
	
	Cosmic strings are one-dimensional defect solutions of field theories\cite{name58}, those can be also regard as cosmologically stretched fundamental strings of String Theory\cite{name59,name60}. Networks of cosmic string are generated in the early universe and expected to exist throughout cosmological history. A kind of SGWB source is cosmic string loops\cite{name5,name6,name7}. Generally, these loops have been generated in abundance throughout the history of cosmology due to the frequent interactions between strings\cite{name23}. After produced by the cosmic strings network, these loops will decay and emit their energy by GWs.
	
	The space-borne detector Laser Space Interferometer Space Antenna (LISA) is sheduled to be launched in the 2030s and aims at detecting gravitational wave around milli-Hz\cite{name61}. TAIJI as another space-borne detector is proposed to be a LISA-like mission and observes the GWs in the same period with LISA\cite{name62}. The joint LISA-TAIJI networks have been studied for the benefits of SGWB detections\cite{name63,name43,name64,name65,name42}.
	
	 In this paper, we seek the capability of joint LISA-TAIJI networks to further limit the tension $G\mu$ (where $G$ is the  gravitational constant) in cosmic string networks.  Analytical approximation is utilized to calculate the energy density of the SGWB\cite{name38} , which is generated by cosmic strings network. For a power-law SWGB we use a special curve named PLS curve\cite{name46} to express the corresponding detectability of the space detectors. For joint LISA-TAIJI, we consider the three different configurations of TAIJI (TAIJIm,TAIJIp,TAIJIc), so the joint network named LISA-TAIJI$x$($x$=m,p,c)\cite{name42}. Comparing the PLS curves of the joint space-borne detectors and SGWB, LISA-TAIJIc has the best sensitivity for SGWB generated by cosmic string networks, around 1mHz, and confine the upper limit of tension $G\mu=O(10^{-17})$.
	 
	 This paper is organized as follows. In Sec.~\ref{II}, we introduce the calculation of SWGB from cosmic strings and its spectral shape. In Sec.~\ref{III}, we introduce the joint networks composed by three configurations of TAIJI and evaluate the equivalent energy density of single detector and joint detectors. In Sec.~\ref{IV}, we compare the PLS curves of the single and joint space detectors with the energy density spectrum of SGWB produced by cosmic string loops, and discuss the detectability of the cosmic string SGWB. We give some summaries in Sec.~\ref{V}
	\section{THE SWGB FROM COSMIC STRINGS}\label{II}
		 The SWGB generated by the evolving universe has been studied in many literatures (cf. \cite{name1,name2,name3,name4,name5,name6,name7,name8,name9,name10,name11,name12,name13,name14,name15,name16,name17,name18,name19,name20,name21,name22}). It is usually quantified as the fraction of the critical density of GW in per logarithmic interval of  frequency
		 \begin{equation}
		 	\Omega_{gw}(t_{0},f)=\frac{8\pi G}{3H_{0}^2}f\frac{\mathrm{d}\rho_{gw}}{\mathrm{d}f}(t_{0},f), \label{1}\tag{2.1}
		 \end{equation}
	 where $H_{0}$ is the Hubble parameter at the current time and $\frac{\mathrm{d}\rho_{gw}}{\mathrm{d}f}(t_{0},f)$ is the energy density of gravitational waves per unit frequency at present. 
	 
	 This paper focuses on SGWBs generated by cosmic strings which have been  well established in \cite{name23,name24,name25,name26,name27,name28,name29,name30,name31,name32,name33}. Cosmic strings are the model of Nambu-Goto(NG) strings. For Eq.\eqref{1} with a given frequency of GW at present, all the GWs emitted by the loops, which are contributive at the frequency, should be integrated throughout the history of the universal evolution to obtain the GW energy at this frequency.

	\subsection{The principle of the GWs energy density calculation in the strings network model}
	Take the redshift of the GW frequency from emission to the present time into account, the energy density of gravitational waves observed today at a particular frequency $f$ is\cite{name23}
	\begin{equation}
		\frac{\mathrm{d}\rho_{gw}}{\mathrm{d}f}(t_{0},f)=G\mu^2 \int_{0}^{t_{0}}\mathrm{d}t(\frac{a(t)}{a_{0}})^3\int_{0}^{\infty}\mathrm{d}ln(t,l)P(\frac{a_{0}}{a(t)}fl)l, \tag{2.2}\label{2}
	\end{equation}
where $G\mu^2$ is dimensionless units of energy and $G\mu$ is the cosmic string tension ($G$ is the gravitational constant, $\mu\approx\eta^2$ is the energy per unit length of string, $\eta$ is the characteristic energy scale), $a(t)$ is the scale factor, at this time its value is taken as $a_{0}$, and $n(l,t)$ is the number density of loops, $P(fl)$ is the average power spectrum of GW, $l$ is the length of loop.
We can use an approximate estimation. Assuming that loops have a periodic behavior in a flat space, the emission should be discrete
\begin{equation}
	\omega_{n}=\frac{2\pi n}{T}, \tag{2.3}\label{3}
\end{equation}
where $T=l/2$ is the oscillation period, $n=1,2,3\ldots$ So we can replace $P(fl)$ by another function $P_{n}$ of the harmonic wave with $n$. A single loop generated by some special events (e.g. Cups Kinks) can emit gravitational wave, which has the power spectrum as\cite{name1,name24,name25,name26}
\begin{equation}
	P_{n}=\frac{\Gamma}{\eta(q)}n^{-q}, \tag{2.4}\label{4}
\end{equation}
where $\eta(q)$ is Riemann zeta function, $\Gamma=\sum_{n=1}^{\infty}P_{n}$ is the total emitted energy. In this paper we take $\Gamma\sim50$ \cite{name27,name28,name29,name30,name31,name32}, and the index $q=5/3,4/3,2$ corresponds to the kinks, cups, kinks-kinks oscillation respectively. The energy density therefore can be expressed to\cite{name27,name30}
\begin{equation}
	\frac{\mathrm{d}\rho_{gw}}{\mathrm{d}f}(t_{0},f)=G\mu^2\sum_{n=1}^{\infty}C_{n}(f)P_{n}, \tag{2.5}\label{5}
\end{equation}
\begin{equation}
	C_{n}=\frac{2n}{f}\int_{0}^{\infty}\frac{\mathrm{d}z}{H(z)(1+z)^4}n(\frac{2n}{(1+z)f},t(z)). \tag{2.6}\label{6}
\end{equation}

In particular, the number of harmonic waves is finite.~However, its total number needs to be well converged for cosmic background results. For standard universe $\Omega_{gw}(t_{0},f)$ will converge in the case of integrating $10^3-10^5$ models. This number depends on the power index $q$, and more models need to be integrated at high frequency\cite{name33}. In our work, the evolution of universe is assumed as a standard $\Lambda CDM$ model whose underlying parameters are $H_{0}=100h~~km/(s\cdot Mpc)$, $h=0.678$, $\Omega_{M}=0.308$, $\Omega_{A}=8.397\times10^{-5}$, $\Omega_{D}=1-\Omega_{A}-\Omega_{M}$.

	In order to calculate the energy density of GW from cosmic strings, it is inevitable to ﬁx the number density function $n\left(l,t\right)$ of their loops, which can be written as\cite{name34,name35,name36,name37,name38}
	\begin{equation}
		n(l,t)=\int_{t_{i}}^{t}\mathrm{d}t'f(l',t')(\frac{a(t')}{a(t)})^3, \tag{2.7}\label{8}
	\end{equation}
where $f(l,t)$ is the production function of the non-self-interacting loops. The above equation can be further specified by numerical simulation of cosmic strings\cite{name27}. So the number densities in the three universe evolution phases are as follows
\begin{align}
	n_{r}(l,t)&=\frac{0.18}{t^{3/2}(l+\Gamma G \mu t)^{5/2}}\Theta(0.1-l/t), \tag{2.8a}\label{9a}\\
	n_{r,m}(l,t)&=\frac{0.18(2\sqrt{\Omega_{A}})^{3/2}(1+z)^3}{(l+\Gamma G \mu t)^{5/2}}\Theta(0.09t_{eq}/t-\Gamma G \mu-l/t), \tag{2.8b}\label{9b}\\
	n_{m}(l,t)&=\frac{0.27-0.45(l/t)^0.31}{t^2(l+\Gamma G \mu t)^2}\Theta(0.18-l/t), \tag{2.8c}\label{9c}
\end{align}
the subscript `r' stands for loops are generated in the radiation era and decay in radiation era; `r,m' means that the loops are generated in the radiation era but emit GWs in the matter era; `m' represents loops are generated and emit GWs in the matter era. $\Theta(x)$ is Heavisdie step function which means through this function we can simply find the limit of loops size in each era, because when $x<0$ the Heaviside function will be zero. In Eq.\eqref{9a} and \eqref{9c} the cutoff values are obtained from the maximum scales that the loops can reach in the corresponding eras. But the cutoff value of Eq.\eqref{9b} is derived from the loops arise in the radiation era and decay in the matter period, which means the loops must exist in a certain size until they are in the matter era.
\subsection{Spectrum of the SGWB from Cosmic Strings loops}
Since the analytical method can calculate the density of loops throughout the history of the Universe and can determine the power spectrum of the loops, we use the analytical method to calculate the cosmic strings in this work, and it has been shown in several related studies that there is not much difference in the conclusions between the analytical and numerical simulation methods\cite{name23}.

From\cite{name38} we can obtain the energy spectrum density of the SGWB in three different phase. The energy spectrum of SGWB with respect to loops generated and decayed in the radiation era is
\begin{equation}
	\Omega_{gw}^{r}(f)=\frac{128}{9}\pi A_{r} \Omega_{A} \frac{G\mu}{\epsilon_{r}}\left[(\frac{f(1+\epsilon_{r})}{B_{r}\Omega_{M}/\Omega_{A}+f})^{3/2}-1\right], \label{10}\tag{2.9}
\end{equation}
here
\begin{equation}
	\epsilon_{r}=\frac{\alpha\xi_{r}}{\Gamma G \mu},~~~A_{r}=\frac{\tilde{c}}{\sqrt{2}}F\frac{\nu_{r}}{\xi_{r}^{3}},~~~B_{r}=\frac{2H_{0}\Omega_{A}^{1/2}}{v_{r}\Gamma G \mu}.\tag{2.10}\label{11}
\end{equation}

The label $r$ indicates the radiation era and  in these equations $\nu_{r}=1/2,\xi_{r}=0.271,v_{r}=0.662,A_{r}=5.4F,F=0.1$, and $\tilde{c}$ is a phenomenological parameter which can be set as $\tilde{c}=0.23\pm0.04$\cite{name35}. Except these parameters there is another parameter $\alpha$ which is used to express the loop size and always be treated as a free constant. 

SGWB from such loops, which arose in the radiation era and decaying in the matter era, has the following energy spectrum density
\begin{small}
	\begin{align}
		&\Omega_{gw}^{r,m}(f)=32\sqrt{3}\pi A_{r}(\Omega_{M}\Omega_{A})^{3/4}H_{0}\frac{A_{r}}{\Gamma}\frac{(\epsilon_{r}+1)^{3/2}}{f^{1/2}\epsilon_{r}}\notag\\
		&\left\{\frac{(\Omega_{M}/\Omega_{A})^{1/4}}{(B_{m}(\frac{\Omega_{M}}{\Omega_{A}})^{1/2}+f)^{1/2}}\left[2+\frac{f}{B_{m}(\Omega_{M}/\Omega_{A})^{1/2}+f}\right]-\frac{1}{(B_{m}+f)^{1/2}}\left[2+\frac{f}{B_{m}+f}\right]\right\},\tag{2.11}\label{12}
	\end{align}
\end{small}
here
\begin{equation}
	B_{m}=\frac{2H_{0}\Omega_{M}^{1/2}}{v_{m}\Gamma G\mu},~~v_{m}=0.583.\tag{2.12}
\end{equation}

In the low frequency region the above spectrum will produce a peak, i.e. when $f\ll B_{r}\Omega_{M}/\Omega_{A}$. 

\begin{figure}
	\includegraphics[width=1\textwidth]{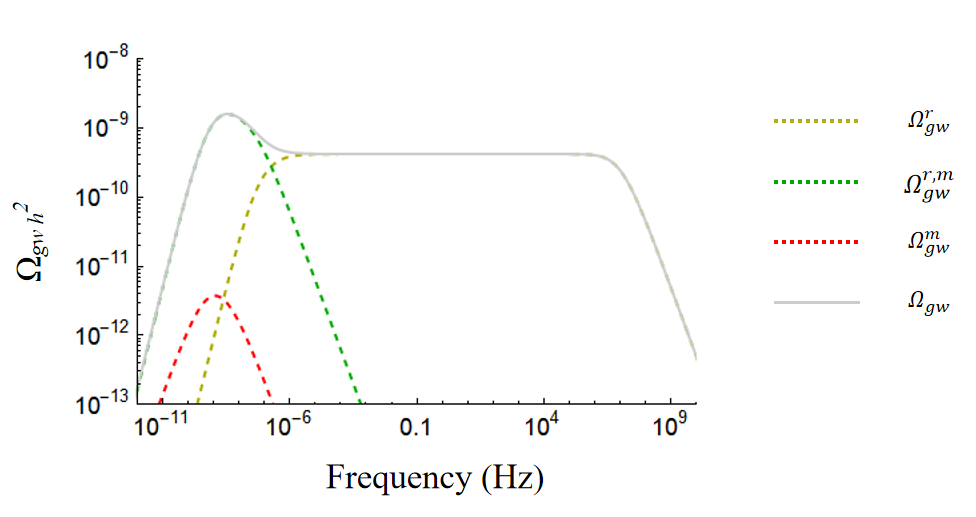}
	\caption{Simulation of the SGWB for cosmic string production.~We choose $G\mu=10^{-10},\alpha=0.1$. The red dashed line expresses the contribution from loops generated and decayed in the matter era, the green dashed line is the contribution from loops generated in the radiation era and decayed in the matter region, the yellow dashed line is the contribution of loops generated and decayed in the radiation region, and the brown curve is the total SGWB.}
	\label{Figure1}
\end{figure}

Loops in the matter era also produce a similar energy spectrum density in the frequency region $f\ll B_{r}\Omega_{M}/\Omega_{A}$
\begin{align}
	\Omega_{gw}^{m}(f)=&54\pi H_{0}\Omega_{M}^{2/3}\frac{A_{m}}{\Gamma}\frac{\epsilon_{m}+1}{\epsilon_{m}}\frac{B_{m}}{f}\notag\\
	&\left\{\frac{2B_{m}+f}{B_{m}(B_{m}+f)}-\frac{2\epsilon_{m}+1}{f\epsilon_{m}(\epsilon_{m}+1)}+\frac{2}{f}\log \left(\frac{\epsilon_{m}+1}{\epsilon_{m}}\frac{B_{m}}{f+B_{m}}\right)\right\},\label{13}\tag{2.13}
\end{align}
here
\begin{equation} A_{m}=\frac{\tilde{c}}{\sqrt{2}}F\frac{\nu_{m}}{\xi_{m}^3},~~~\epsilon_{m}=\frac{\alpha\xi_{m}}{\Gamma G\mu},\tag{2.14}
\end{equation}
and we get $\nu_{m}=2/3,\xi_{m}=0.625,v_{m}=0.583,A_{m}=0.39F$. $\Omega_{gw}^{r,m}$ should be dominant when $\alpha\gg\Gamma G\mu$ , and its dominance decreases with the reduction of $\alpha$ until $\Omega_{gw}^{r,m}\sim\alpha^{1/2}$. When $\alpha$ is sufficiently small, $\Omega_{gw}^{m}$ will dominate in the low frequency region. The spectra of the three eras are shown in Figure \ref{Figure1}.

For $\alpha\geq\Gamma G\mu$ and $f<3.5\times10^{10}/(1+\epsilon_{r})Hz$, SGWB can be well approximated by the following form\cite{name38}
\begin{equation}
	\Omega_{gw}(f)=\Omega_{gw}^{r}(f)+\Omega_{gw}^{r,m}(f)+\Omega_{gw}^{m}(f). \tag{2.15}\label{14}
\end{equation}

The spectra of SGWB all exhibit in the Figure \ref{Figure1}. For small strings, i.e. $\alpha<\Gamma G \mu$, the loops will decay rapidly so that there are not loops can be produced in radiation era and exist in matter era. Therefore, the SGWB formulation would be further simplified as\cite{name20,name38}
\begin{equation}
	\Omega_{gw}(f)=\frac{64}{3}\pi G\mu\Omega_{A}A_{r}+54\pi\frac{H_{0}\Omega_{M}^{3/2}}{\epsilon_{m}\Gamma}\frac{A_{m}}{f}\left[1-\frac{B_{m}}{\epsilon_{m}f}\right].\tag{2.16}\label{15}
\end{equation}

\begin{figure}
	\centering
	\includegraphics[width=1\textwidth]{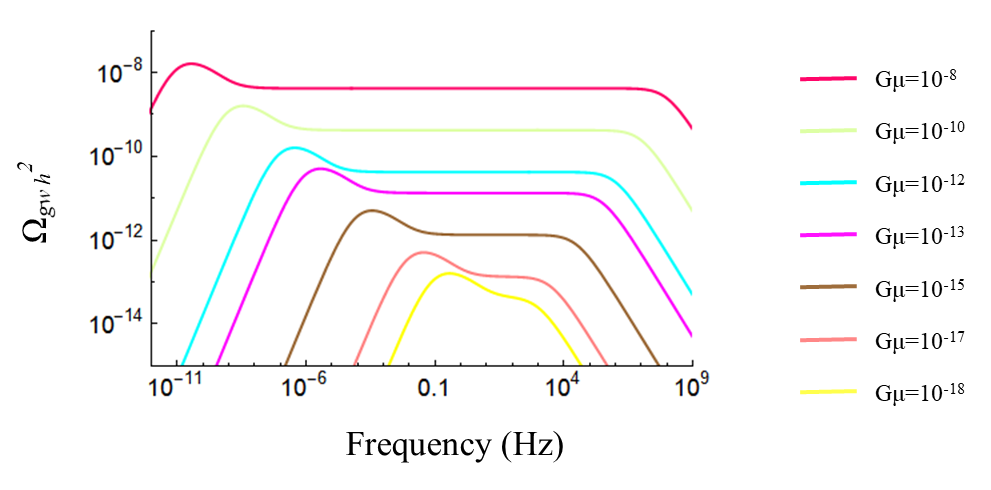}
	\caption{The spectrum of SGWB varying with $G\mu$ and the free parameter constant $\alpha= 0.1$.}
	\label{Figure2}
\end{figure}

Similarly if we take different values of $\alpha$ and $G\mu$ we will get different curves, in this simulation we chosen the value of $\alpha$ as 0.1, so the spectrum of SGWB varying with $G\mu$ is shown in Figure \ref{Figure2}.

\section{Detectability of SGWB from cosmic strings in the space detectors}\label{III}
The LISA and TAIJI detectors are designed to detect gravitational wave signals in space, and as the mission of both detectors are of the same duration, we are looking forward to exploring the SGWB from cosmic string through a joint detection between the detectors.

LISA launched by EPA $\left(European~Space~Agency\right)$, consists of three spacecrafts separated by trailing behind the Earth while it orbits the Sun. These three spacecrafts relay laser beams back and forth in the channel between different spacecraf and the signals are combined to search for GWs around 1mHz. TAIJI as a LISA-like detector share the same geometry and path, but its arm length is 3 million kilometers and ahead the Earth.

For LISA-like detectors Time-delay interferometry (TDI) can suppress the laser frequency noise, which is to combine multiple time-shifted interferometric links and obtain an equivalent equal path for two interferometric laser beams\cite{name42}.

Using the symmetry of the system about a single LISA-like triangular unit that can compose three data channels $A,E$ and $T$ without correlated noises\cite{name43,name47}. Due to the $T$ channel has a negligible effect on the frequency range of our study, we only consider $A$ and $E$ channels.
\subsection{The noise in each detector}
The equivalent energy density can characterize the sensitivity of the detector to SGWB. For a LISA-like mission it could be evaluated as\cite{name42}
\begin{equation}
	\Omega_{mission}(f)=\frac{4\pi^2f^3}{3H_{0}^2}\left(\sum_{i=A,E}\frac{R_{i}(f)}{N_{i}(f)}\right)^{-1},\tag{3.1}\label{3.1}
\end{equation}
where $R_{i}(f)$ is the response function of the relevant channel, $N_{i}(f)$ is the noise spectrum. According to ref.\cite{name39},
\begin{align}
	R_{A,E}&\cong\frac{9}{20}\left|W\right|^2\left[1+\left(\frac{f}{4f_{*}/3}\right)^2\right]^{-1},\tag{3.2a}\\
	W&=1-e^{-2if/f_{*}},\tag{3.2b}
\end{align}
where $f_{*}$ is the characteristic frequency of a single detector and $f_{*}=c/\left(2\pi L\right)$.

 The noise spectrum $N_{i}(f)$ can be found in LISA Science Requirements Document\cite{name40}. And the noise of $A,E$ channels is\cite{name41} 
 \begin{equation}
 	N_{A,E}\cong\left|W\right|^2\left(6N_{o}(f)+24N_{a}(f)\right).\tag{3.3}\label{3.3}
 \end{equation}
 
 The primary noises in this idealized model are acceleration noise $N_{a}$ and optical path perturbation noise $N_{o}$ which can be expressed as follows
		\begin{equation}
			N_{a}=\frac{N_{I}}{4\left(2\pi f\right)^4},~~~~~N_{o}=N_{II},\tag{3.4}
		\end{equation}
	where
	\begin{align}
		N_{I}&=4\left(\sqrt{(\delta a)^2}/L\right)^2\left(1+(f_{1}/f)^2\right)\notag\\
		&=5.76\times10^{-48}\times(1+(f_{1}/f)^2)~~s^{-4}Hz^{-1},\tag{3.5}\label{3.5}\\
		N_{II}&=\left(\sqrt{(\delta x)^2}/L\right)^2=3.6\times10^{-41}~Hz^{-1},\tag{3.6}\label{3.6}
	\end{align}
in which $f_{1}=0.4~~mHz$, and the $\mathbf{rms}$ magnitudes are
\begin{equation}
	\sqrt{\left(\delta a\right)^2}=3\times10^{-15} ~~m/s^2,~~~\sqrt{\left(\delta x\right)^2}=1.5\times10^{-11}~~ m.\tag{3.7}\label{3.7}
	\end{equation}

$L=2.5\times10^6$km is the arm length of LISA, and $L=3.0\times10^6$km is used to calculate TAIJI (considering the arm length of each detector is constant in the follow text). TAIJI has the same acceleration noise and optical path perturbations as LISA \cite{name39}.

Combining equations \eqref{3.1}—\eqref{3.7}, the equivalent energy density equation of a single detector can be expressed as
\begin{equation}
	\Omega_{mission}(f)=\frac{20\pi^2f^3}{3H_{0}^2}\left(\frac{5.76\times10^{-48}(f^2+f_{1}^2)}{16\pi^4f^6}+3.6\times10^{-41}\right)R(f),\tag{3.8}\label{3.8}
\end{equation}

where
\begin{align}
	R(f)&=1+\left(\frac{f}{f_{2}}\right)^2,\tag{3.9}\\
	f_{2}&=\frac{4f_{*}}{3}.\tag{3.10}
\end{align}

\subsection{Cross-Correlation analysis}
For Cross-Correlation, i.e. LISA-TAIJI network, we adopt LISA and three alternative TAIJI orbital deployment to construct the network\cite{name42}
\begin{enumerate}[a)]
	\item LISA, trailing the Earth by $\sim 20^{\circ}$ and its formation plane have an inclination angle respect to the ecliptic plane about $\sim+60^{\circ}$.
	\item TAIJIm, leading the Earth by $\sim 20^{\circ}$, with a $\sim-60^{\circ}$ inclination.
	\item TAIJIp, leading the Earth by $\sim 20^{\circ}$, with a $\sim+60^{\circ}$ inclination.
	\item TAIJIc, trailing the Earth by $\sim 20^{\circ}$, is coplanar with LISA.
\end{enumerate}

The equivalent energy density of LISA-TAIJI network can be evaluated as\cite{name42},
\begin{equation}
	\Omega_{cross}(f)=\frac{4\pi^2f^3}{3H_{0}^2}\left(\sum_{i,j=A,E,T}\frac{\left|\gamma_{ij}(f)\right|^2}{S_{n,i}^{LISA}(f)S_{n,j}^{TAIJI}(f)}\right)^{(-1/2)},\tag{3.11}\label{3.11}
\end{equation}
the overlap reduction function $\gamma_{ij}(f)$ can be written as\cite{name43}
\begin{equation}
	\gamma_{ij}(f)=\Gamma_{abcd}D_{i,ab}D_{j,cd},\tag{3.12}\label{3.12}
\end{equation}
moreover, the tensor $\Gamma_{abcd}$ can be written as a formula of Kronecker's delta and unit vector $m_{i}$
\begin{equation}
	\Gamma_{abcd}=b_{0}\delta_{ac}\delta_{bc}+b_{1}\delta_{ac}m_{b}m_{d}+b_{2}m_{a}m_{b}m_{c}m_{d},\tag{3.13}\label{3.13}
\end{equation}
the subscript $abcd$ is the four arms of two effective L-shaped interferometers $A$ and $E$, and the coefficient $b_{0},b_{1},b_{2}$ are given by the spherical Bessel function $j_{l}=j_{l}(y)$, $y=2\pi fd/c$, where $d$ is the separation distance between the two detectors
\begin{equation}
	\label{3.14}\tag{3.14}
	\begin{split}
		b_{0}(y)&=2\left(j_{0}-\frac{10}{7}j_{2}+\frac{1}{14}j_{4}\right),\\
		b_{1}(y)&=4\left(\frac{15}{7}j_{2}-\frac{5}{14}j_{4}\right),\\
		b_{2}(y)&=\frac{5}{2}j_{4}.\\
	\end{split}
\end{equation}

$D_{i,ab}$ and $D_{j,cd}$ are the tensors of the two detectors. Considering the L-shaped interferometer $A$ for a single triangle unit $X$ and using the orthonormal unit vector $\left(\mathbf{m_{a},\mathbf{m_{b}}}\right)$, then the detector tensor for the A channel can be
\begin{equation}
	D_{A,ab}=\left(\mathbf{m_{a}}\bigotimes \mathbf{m_{a}}-\mathbf{m_{b}}\bigotimes \mathbf{m_{b}}\right)/2.\tag{3.15}\label{3.15}
\end{equation}

The $A$ and $E$ channels can be effectively regarded as two L-shaped interferometers with an offset angle of $45^{\circ}$\cite{name43}. Therefore, the detector tensor for the $E$ channel is
\begin{equation}
	D_{A,ab}=\left(\mathbf{m_{a}}\bigotimes \mathbf{m_{b}}+\mathbf{m_{a}}\bigotimes \mathbf{m_{b}}\right)/2.\tag{3.16}\label{3.16}
\end{equation}

$\mathbf{m_{a}},\mathbf{m_{b}}$ are the orientation of the two arms about a L-shaped interferometer $(\mathbf{m_{a}}\cdot \mathbf{m_{b}}=0)$. Ignoring the effect of channel T when calculating the equivalent energy density of the joint network, since its effect is not significant\cite{name43}. That is, the joint network we are studying consists of four part, namely $AA,AE,EA,EE$ (without considering the beam pattern function). The noise models used in the joint network in\cite{name44} are expressed as follows
\begin{equation}
	S_{n,i}^{LISA}(f)=\frac{4}{3R_{L}^2}\left[P_{o1}+2\left[1+\cos{(f/f_{*})}^2\right]\frac{P_{a1}}{(2\pi f)^4}\right]\times\left[1+0.6(f/f_{*})^2\right],\tag{3.17}\label{3.17}
\end{equation}
\begin{equation}
	S_{n,j}^{TAIJI}(f)=\frac{4}{3R_{T}^2}\left[0.8^2P_{o1}+2\left[1+\cos{(f/f_{*})}^2\right]\frac{P_{a1}}{(2\pi f)^4}\right]\times\left[1+0.6(f/f_{*})^2\right],\tag{3.18}\label{3.18}
\end{equation}
where
\begin{align}
	P_{a1}&=9.0\times10^{-30}\left[1+(4\times10^{-4}/f)^2\right]\notag\\
	&\times\left(1+\left[f/(8\times10^{-3})\right]^4\right)~~m^2s^{-4}/Hz,\tag{3.19}\\
	P_{o1}&=2.25\times10^{-22}\left[1+\left(2\times10^{-3}/f\right)^4\right]~~Hz^{-1},\tag{3.20}
\end{align}

Note that for LISA and TAIJI, the characteristic frequency $f_{*}$ is not the same, and $f_{*}$ is related to the arm length of the detector and $R_{L}=2.5\times10^6$km, $R_{T}=3.0\times 10^{6}$km. Since the noises of the A and E channels are the same, we can define the total response function as
\begin{equation}
	Y(f)=\gamma_{AA}^2+\gamma_{AE}^2+\gamma_{EA}^2+\gamma_{EE}^2.\tag{3.21}\label{3.20}
\end{equation}

By combining Eq.\eqref{3.12}--\eqref{3.16} and Eq.\eqref{3.20} it can obtain that
\begin{equation}
	Y(f)=\sum_{e=0}^{2}\sum_{f=0}^{2}b_{e}(y)b_{f}(y)X_{ef},\tag{3.22}\label{3.21}
\end{equation}
$X_{ef}$ can be obtained by the tensor of the detector and the unit orientation vector $m_{i}$. For instance
\begin{equation}
	X_{02}=\sum_{i}^{AE}\sum_{j}^{AE}\left(\delta_{ac}\delta_{bd}D_{i,ab}D_{j,cd}\right)\left(m_{r}m_{s}m_{t}m_{u}D_{i,rs}D_{j,tu}\right).\tag{3.23}
\end{equation}

 The total response function and tensor factors are rotationally invariant\cite{name43}, and the tensor factor $X_{ef}$ for acquiring $Y(f)$ is the key point. For calculating $X_{ef}$ there are three cosines
\begin{equation}
	c_{l}=\hat{e_{l}}\cdot \mathbf{m},~~~~~c_{tx}=\hat{e_{tx}}\cdot \mathbf{m},~~~~~c_{ltx}=\hat{e_{l}}\cdot\hat{e_{tx}},\tag{3.24}\label{3.23}
\end{equation}
$\hat{e_{l}},\hat{e_{tx}}$ are the normalized unit vectors and $\mathbf{m}$ is the unit direction vector of the detector LISA and TAIJI(m,p,c). The subscript $l$ represents LISA and $tx$ indicates the different orbital deployments of TAIJI $x=m,p,c$. According to\cite{name43}, $Y(f)$ can be simplified as 
\begin{equation}
	Y=\sum_{e=0}^{2}\sum_{f=0}^{2}b_{e}(y)b_{f}(y)X_{ef}(c_{l},c_{tx},c_{ltx}).\tag{3.25}\label{3.25}
\end{equation}
and the detailed expansion of $X_{ef}$ is given by Eq.(40)--(45) in the Ref.\cite{name43}.

The values of the unit vectors for TAIJI(m,p,c) and  normalized unit vectors are as follows
\begin{enumerate}[A.]
	\item For LISA-TAIJIm, the orbital and structural design gives $d_{m}=1AU\times2\sin{20^{\circ}}$, and the directional vector $\mathbf{m}=\left(0,1,0\right)$. The normalized unit vector of the detector for this joint network is
	\begin{equation}
		\hat{e_{tm}}=\left(-\frac{\sqrt{3}}{2}\cos{20^{\circ}},-\frac{\sqrt{3}}{2}\sin{20^{\circ}},-\frac{1}{2}\right),\tag{3.26}
	\end{equation}
\item For LISA-TAIJIp, the orbital and structural design gives $d_{p}=1AU\times2\sin{20^{\circ}}$, and the directional vector $\mathbf{m}=\left(0,1,0\right)$. The normalized unit vector of the detector for this joint network is
\begin{equation}
	\hat{e_{tp}}=\left(-\frac{\sqrt{3}}{2}\cos{20^{\circ}},-\frac{\sqrt{3}}{2}\sin{20^{\circ}},\frac{1}{2}\right),\tag{3.27}
\end{equation}
\item For LISA-TAIJIm, the orbital and structural design gives $d_{m}=0$, and the directional vector $\mathbf{m}=\left(0,0,1\right)$. The normalized unit vector of the detector for this joint network is
\begin{equation}
	\hat{e_{tc}}=\left(-\frac{\sqrt{3}}{2}\cos{20^{\circ}},\frac{\sqrt{3}}{2}\sin{20^{\circ}},\frac{1}{2}\right).\tag{3.28}
\end{equation}
\end{enumerate}

Combining Eq.\eqref{3.11}——\eqref{3.25} and ignoring the effect of the T channel we can obtain the equivalent energy density of the joint LISA-TAIJI$x$ network, since the noises in the A and E channels are the same,  we can express them as such and simplify the subscript $ni,nj$ to $n$
\begin{equation}
	\Omega_{crossx}(f)=\frac{4\pi^2f^3}{3H_{0}^2}\left(\frac{\sum_{e=0}^{2}\sum_{f=0}^{2}b_{e}(y)b_{f}(y)X_{ef}(c_{l},c_{tx},c_{ltx})}{S_{n}^{LISA}(f)S_{n}^{TAIJI}(f)}\right)^{-1/2}.\tag{3.29}
\end{equation}

\begin{figure}
	\centering
	\includegraphics[width=1\textwidth]{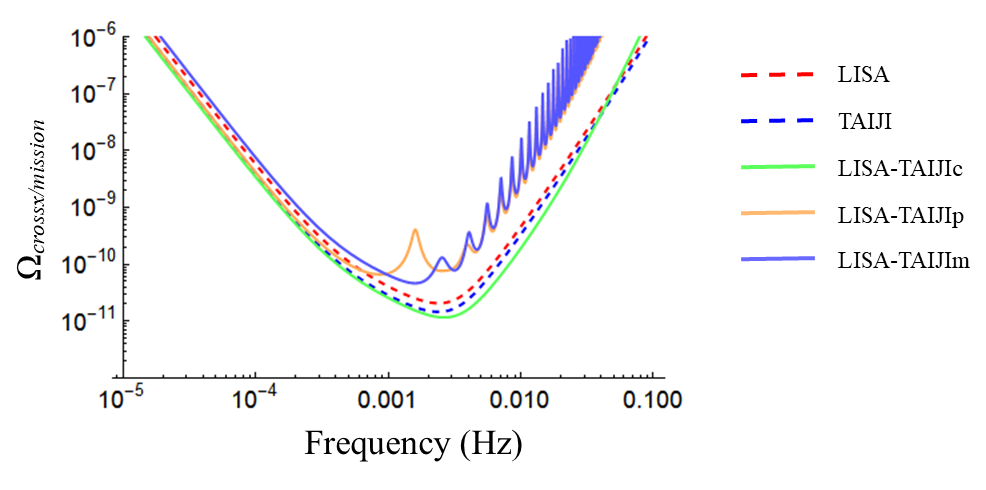}
	\caption{The spectrum of equivalent energy density. Different curves express different sensitivities, and it can be seen that the sensitivity of LISA-TAIJIc is the best in all designs when the effect of the T-channel is not accounted for, and it can show good performance in the low-frequency region for all joint observations.}
	\label{Figure3}
\end{figure}

The subscript $crossx$ can be taken as $crossm,crossp$ and $crossc$, denoting the joint observation of LISA and TAIJI(m,p,c) respectively. The full spectrum of equivalent energy density is shown in Figure \ref{Figure3}.

\section{Detection of SGWB from cosmic strings by the space detectors}\label{IV}
Generally the Power-Law sensitivity (PLS) is used to express the detectability of a power-law SGWB in a GW detector\cite{name46,name42}. Through comparing with the spectrum of the power-law SGWB energy density, the PLS curve can illuminate whether the detector is able to search the SGWB or not. There are two physical quantities are important for calculating the PLS, which are (1) the signal to noise threshold $\rho_{t}$, (2) the observation time $T_{ob}$.

Based on the $\rho_{t}$ and $T_{ob}$, the detector PLS can be calculated as
\begin{equation}
	\Omega_{\kappa}=\frac{\rho_{t}}{\sqrt{2}T_{ob}}\left(\int_{0}^{f_{max}}\frac{(f/f_{ref})^{2\kappa}}{\Omega_{mission}(f)^2}\right)^{-1/2},\tag{4.1}\label{4.1}
\end{equation}
where the subscript $mission$ in Eq.\eqref{4.1} represents a single detector or a joint detector, which can be replaced by $crossx$ when LISA-TAIJI(m,p,c). For a set of power-law indices e.g., $\kappa\in\{-8,-7,\ldots,7,8\}$. The reference frequency $f_{ref}$ can be chosen arbitrarily and will not impact on the PLS curve\cite{name46}. For each value of $\kappa$ and different $f_{ref}$, the $\Omega_{\kappa}$ can be derived. Then the power-law sensitivity $\Omega_{PLS}$ can be given by the following equation
\begin{equation}
	\Omega_{PLS}(f)=\max_{\kappa}\left[\Omega_{\kappa}\left(\frac{f}{f_{ref}}\right)^{\kappa}\right].\tag{4.2}
\end{equation}

\begin{figure}
	\centering
	\includegraphics[width=1\textwidth]{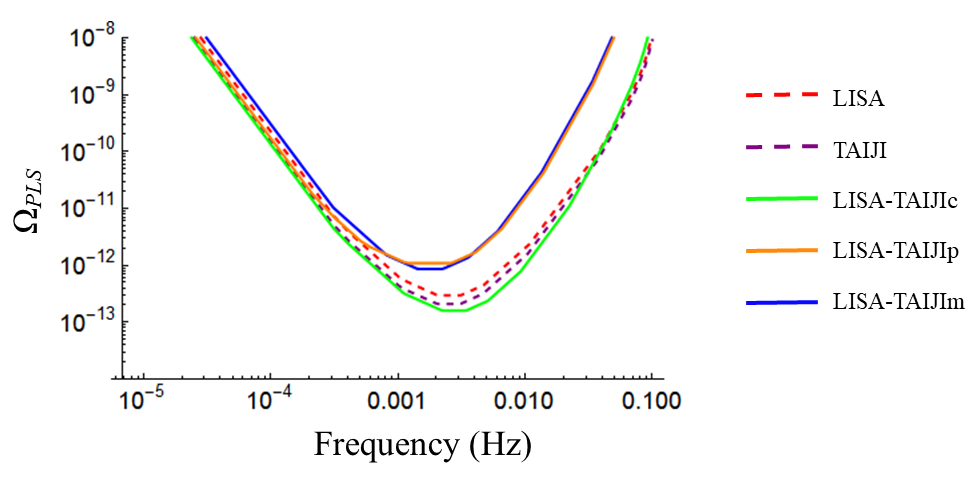}
	\caption{The PLS curves for different detection methods. In $\rho_{t}=10$ and $T_{ob}=3~years$, the PLS plots of different detection methods, show that LISA-TAIJIc still has the best sensitivity without considering the effect of the T-channel, and LISA-TAIJIc consistently shows the best detection capability under the joint network.}
	\label{Figure4}
\end{figure}

Figure \ref{Figure4} shows the PLS curves for different detection methods such as the individual or joint detectors. The observation time of three years is chosen since the LISA and TAIJI will be in co-operational mode for three years during their mission, so $T_{ob}=3$ years. Here $\rho_{t}=10$.

\begin{figure}
	\centering
	\includegraphics[width=1\textwidth]{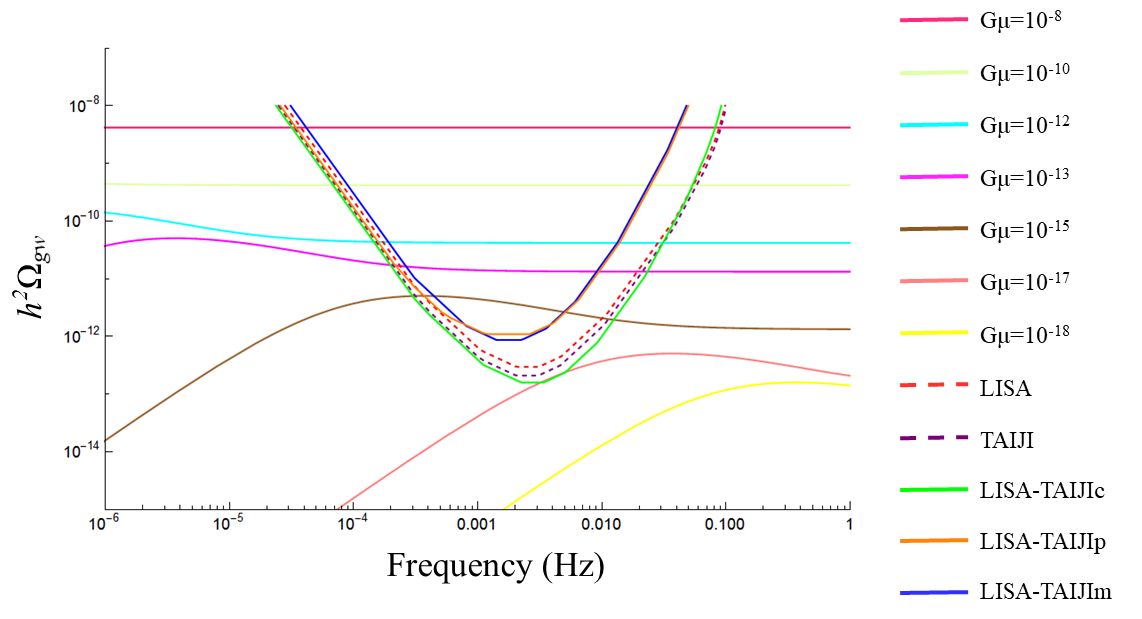}
	\caption{The result of combining $\Omega_{gw}$ and PLS curves. It can be seen that a single space detector can detect the SGWB signal at $G\mu=10^{-15}$, but not at $G\mu=10^{-17}$. However, for LISA-TAIJIc the combined detection can still achieve a SNR$>$10 for SGWB in the tension $G\mu=10^{-17}$ around 1 mHz.}
	\label{Figure5}
\end{figure}

We will use the PLS diagram to express the possibility of observing the  gravitational wave signals generated by cosmic strings, i.e. comparing the $\Omega_{gw}$ in Figure \ref{Figure2} with the PLS in Figure \ref{Figure4}. Figure \ref{Figure5} can help us to judge whether the SGWB signals can be detected and whether the joint detection can further limit the tension $G\mu$.

From the results, it can be found that LISA-TAIJIc can detect SGWB generated by cosmic strings at $\alpha=0.1$ and the tension $G\mu=10^{-17}$.~Also within the PLS diagram, LISA-TAIJIc still shows the most favorable sensitivity, while LISA-TAIJIp still has a better capability of detection than the individual detectors in the low frequency region.

\section{Conclusion}\label{V}
In this paper, we analyze the detection ability of the single and joint space detectors for the SGWB generated by cosmic string loops. By comparing the PLS curves with energy density spectrum of SGWB, all the single and joint detectors are capable for capturing the SGWB with $G\mu\geq10^{-15}$. Among them, the sensitivities of LISA-TAIJIp, LISA-TAIJIm become less than the single detectors (i.e., LISA and TAIJI), which is consist with the results in Ref.\cite{name42},~the LISA-TAIJIc detector network has the best sensitivity for detecting the SGWB from cosmic string loops. It might be able to detect SGWB with loop size $\alpha=0.1$ and tension $G\mu=10^{-17}$. In out results, the sensitivity of LISA-TAIJIc surpasses that of TAIJI, which is slightly different from the result in Figure 2 of Ref.\cite{name42}. That is because we use an analytical approximation to calculate the SGWB generated by cosmic string loops, yielding some divergences from the numerical results. And in the calculation for the spectrum of equivalent energy density and PLS curves, we ignore the effect of T channel and use an analytical method rather than a numerical simulation.

This work provides a future scientific goal for probing SGWB during LISA and TAIJI operations. As another important GW space-borne detector, TIANQIN can be joined into the detection network. Such studies will be an important issue in our follow-up research, that is potential to further restrict the parameters of SGWB generated by cosmic string loops or other theoretical models.

\section*{Acknowledgments}
We are thankful to Dr. Gang Wang for many helpful discussions. This work was supported by the National Key Research and Development Program of China (Grant No. 2021YFC2203004), the National Natural Science Foundation of China (Grant No. 11873001, 12147102)
	
	\bibliographystyle{unsrt}
	\bibliography{ref}
\end{document}